\documentclass[aip,apl,preprint,superscriptaddress]{revtex4}
\pdfoutput=1
\usepackage{graphicx}%

\begin{document}
\title{Controlling the growth of Bi(110) and Bi(111) films on an 
insulating substrate}
\author{Maciej \surname{Jankowski}}
\affiliation{ESRF-The European Synchrotron,71 Avenue des Martyrs, 38000 
Grenoble, France}
\author{Daniel \surname{Kami\'{n}ski}}
\affiliation{Department of Chemistry, University of Life Sciences in Lublin, 
20-950, Poland}
\author{Kurt Vergeer}
\affiliation{University of Twente, Inorganic Materials Science, MESA$^+$ 
Institute for Nanotechnology, P.O. Box 217, NL-7500AE Enschede, The Netherlands}
\author{Marta \surname{Mirolo}}
\affiliation{ESRF-The European Synchrotron,71 Avenue des Martyrs, 38000 
Grenoble, France}
\author{Francesco \surname{Carla}}
\affiliation{ESRF-The European Synchrotron,71 Avenue des Martyrs, 38000 
Grenoble, France}
\author{Guus \surname{Rijnders}}
\affiliation{University of Twente, Inorganic Materials Science, MESA$^+$ 
Institute for Nanotechnology, P.O. Box 217, NL-7500AE Enschede, The Netherlands}
\author{Tjeerd R.J. \surname{Bollmann}}
\affiliation{University of Twente, Inorganic Materials Science, MESA$^+$ 
Institute for Nanotechnology, P.O. Box 217, NL-7500AE Enschede, The Netherlands}


\begin{abstract}
Here we demonstrate the controlled growth of Bi(110) and Bi(111) films 
on 
an (insulating) $\alpha$-Al$_2$O$_3$(0001) substrate by 
surface X-ray diffraction and X-ray reflectivity using synchrotron radiation.
At temperatures as low as 40 K, unanticipated pseudo-cubic Bi(110) films are 
grown having a thickness ranging from a few to tens of nanometers. The 
roughness at the film-vacuum as well as at the film-substrate 
interface, can be reduced by mild heating, where a crystallographic orientation 
transition of Bi(110) 
towards Bi(111) is observed at 400~K. From 450~K onwards high quality and 
ultrasmooth Bi(111) films are formed. Growth around the transition 
temperature results in the growth of competing Bi(110) and Bi(111) thin film 
domains.       

\end{abstract}

\pacs{} 
%
\maketitle


\section{Introduction}
Nanostructured ultrathin Bi films have recently attracted  a lot of 
interest as they reveal exotic magneto-electronic properties making them 
appealing materials for spintronic applications
\cite{Hofmann2006, Koroteev2008, Xiao2012_MJ, Yang2012, Ando2013, Sabater2013, 
Aitani2014_MJ, 
Drozdov2014_MJ, Aguilera2015, Miao2015, Du2016_MJ, Reis2016}. Especially the 
spin-momentum locked 
surface states of topological insulating Bi films \cite{Song2015, 
Lu2015_MJ, Yao2016, 
Bian2016}, make them very attractive 
candidates for spintronic devices. 
To develop and optimize topological insulators (TIs) towards applications, thin 
films of high quality 
are a necessity, as otherwise the exotic electronic properties are hampered by 
bulk conduction \cite{Liu2011, Aitani2014_MJ, Lima2015}. To minimize the 
contribution of the substrate \cite{Miao2015}, an atomically well defined 
insulating substrate, providing an infinite potential well barrier, is essential 
for both future electronic applications as well as to get a deeper understanding 
on the controllability of Bi growth. This choice of substrate is also very 
beneficial for practical applications, as the interface between film and 
insulating substrate, expected to reveal topological states, will also be 
protected from influencing oxidation effects arising from ambient exposure in 
technological applications \cite{Tabor2011}.
The 
growth of Bi has been extensively 
studied on Si(111) \cite{Tanaka1999, Nagao2000, Yaginuma2003, Nagao2004, 
Kammler2005, Moenig2005, Nagao2005, Yaginuma2007, Luekermann2014, Kokubo2015} 
and HOPG \cite{Scott2005, McCarthy2010,Kowalczyk2011, Song2015} as well as 
other surfaces \cite{Jeffrey2006, Kim2006, Hattab2008, Payer2008, Zhang2009, 
Bobaru2012, Xiao2012_MJ, Yang2012, Reis2016}, resulting in fabrication of films 
with a range of 
different morphologies, orientations, and strain.
The fabrication of Bi films has attracted considerable interest in recent 
years, as their controlled growth, with focus on morphology and 
crystallographic orientation, on semiconductor and oxide surfaces is not a 
trivial task. It is well-known that metals on semiconductors and oxides  
usually show 3D growth modes \cite{Campbell1997} instead of atomically smooth 
(2D) films. However, this problem can be overcome by use of 
deposition at low temperatures \cite{Ernst1993, Campbell1994} or
surfactant-mediated growth \cite{Rosenfeld1993, Zhang1997}, as it
modifies the film kinetics.
%


In this study we demonstrate by surface X-ray diffraction (SXRD) the controlled 
growth of thin Bi(110) and Bi(111) films (the index used throughout this paper 
refers to the rhombohedral system) on such an insulating substrate: atomically 
smooth 
insulating sapphire ($\alpha$-Al$_2$O$_3$(0001)) having a lattice mismatch of 
4.6\% with Bi(111), so large that thermal mismatch might be ignored. The 
preparation of pseudo-cubic (110)-oriented Bi films, a 
rather exotic orientation, is a difficult task \cite{Wu2008}. At temperatures 
as low as 40~K, we are able to slow down kinetics resulting in a high 
nucleation density of Bi islands and thereby 
controlling the growth of Bi towards smooth Bi(110) films, stable up to 400~K. 
By annealing the Bi(110) films beyond this temperature, they can be transformed 
towards stable Bi(111) films. For films grown around RT, a competition 
between (110) and (111) thin film domains is observed.

\section{Experimental}
For the surface X-ray diffraction (SXRD) experiments described here, we used hat 
shaped $\alpha$-Al$_2$O$_3$(0001) single crystals with a miscut of 
$<$0.2$^\circ$. Prior to annealing for 12 hours in a tube furnace at 1323~K 
using 
an O$_2$ flow of 150~l/h, the samples have been ultrasonically degreased in 
acetone and ethanol. The samples were then initially inspected by tapping mode 
atomic force microscopy (TM-AFM) for their stepheight (0.21~nm between two 
adjacent 
oxygen planes) and terrace width ($\sim$300~nm,) and X-ray photoelectron 
spectroscopy (XPS) to verify the surface cleanliness where only minor traces of 
C and Ca were found, see Supplemental Material. After insertion into the UHV 
system of the surface diffraction beamline 
ID03/ESRF (Grenoble, France) \cite{Balmes2009} with a base pressure below 
~1$\times$10$^{-10}$~mbar, the sample was cleaned by mild 700~eV Ar$^+$ 
sputtering at $p(Ar)=3\times10^{-6}$~mbar and subsequent annealing to ~1200~K in 
an O$_2$ background pressure of $1\times10^{-6}$~mbar cycles, where we 
monitored 
the sample quality by Auger electron spectroscopy (AES), see 
%
Supplemental 
Material.
Bi was deposited at a typical deposition rate of 1.3\AA\ 
per minute 
from a Mo crucible mounted inside an electron-beam evaporator (Omicron EFM-3). 
According to the bulk phase diagram, Bi and sapphire are immiscible in the 
bulk \cite{Okamoto1991}. The surface X-ray diffraction (SXRD) experiments were 
performed using a monochromatic synchrotron X-ray beam at 24~keV and a 
{MAXIPIX} 
detector\cite{Ponchut2011} with 512$\times$512 pixels. For data integration and 
the creation of reciprocal space maps from the 2D detector frames we used 
the BINoculars software package \cite{Roobol2015}. All reciprocal space 
positions are 
given in (h,k,l) measured in reciprocal lattice units (r.l.u.) of the hexagonal 
substrate (0001) surface lattice. Bragg peaks of the thin Bi films are labeled 
by 
their 
conventional rhombohedral Miller indices \cite{Hofmann2006}. X\mbox{-}ray 
reflectivity (XRR) curves have been fitted using the GenX software package 
\cite{GenX}.
\section{Results}



\begin{figure}
\includegraphics[width=0.8\textwidth]{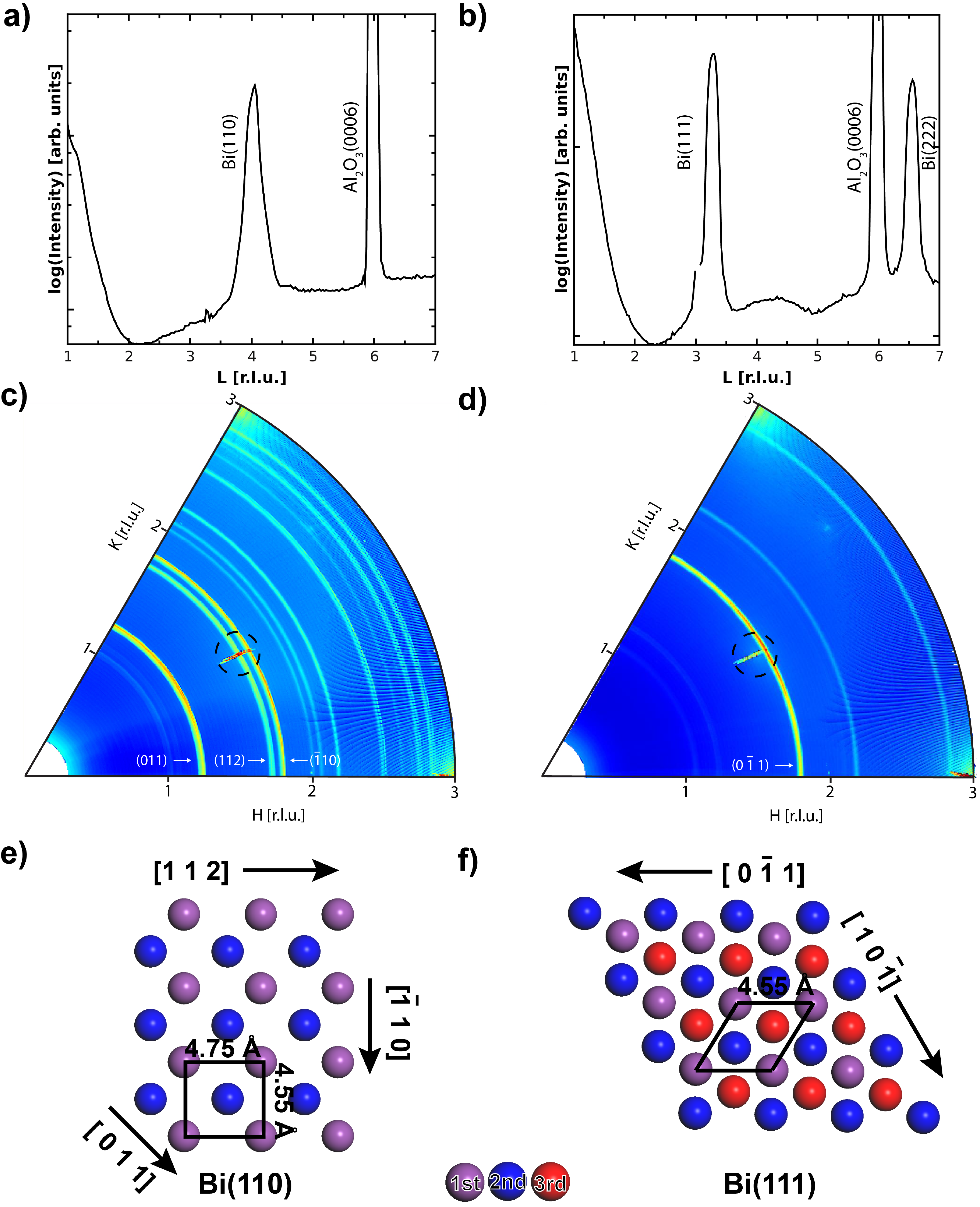}
\caption{(color online) (00) CTR scan for a 20~nm thick Bi(110) film 
deposited at 40~K revealing the Bi(110) Bragg peak at L=4  (a) and a 20~nm thick 
Bi(111) film revealing the Bi(111) Bragg peak at L=3.3 after annealing to 400 K 
(b). The corresponding reciprocal space maps for the 20~nm thick Bi(110) 
measured at L=0.3 at RT (c) and for the annealed
Bi(111) (d) film. The substrate (11) CTR is marked by dashed circles. Next 
to the diffraction rings, resulting from the 
rotational disordered domains, the corresponding miller indices of their
crystallographic planes are denoted. 
(e) A ball model of the pseudo-cubic Bi(110) 
surface. (f) A ball model of the hexagonal Bi(111) surface. Atoms in the 1st, 
2nd and 3rd layer 
are marked by purple, blue and red colors, respectively.}
\label{fig:1}
\end{figure}

In order to determine the surface structure and morphology of the thin Bi 
films grown in-situ, we make use of XRR scans, crystal truncation rod (CTR) 
scans and reciprocal space maps 
determined by SXRD. 
The out-of-plane (electronic) density profile measured by 
XRR provides information on film layer density, film thickness and interface 
roughness. 
The measured (00) crystal truncation rod (CTR) provides information on the 
out-of-plane  crystallographic orientation of the film.
To be sensitive to the in-plane registry we record reciprocal 
spacemaps (at constant L=0.3). In Fig.~\ref{fig:1}(a-b) and (c-d) we show the 
(00) CTR and 
reciprocal space map of thin Bi films grown on the sapphire substrate. A sharp 
pronounced (0006) Bragg peak in (a-b) corresponds to the out-of-plane 
interlayer distance of the sapphire (0001) surface unit cell. 
Upon growth of a 20~nm thick film at 40~K, a Bragg-peak is found at 
L=4 corresponding to the 3.25\AA\ interlayer distance of Bi(110) 
\cite{Hofmann2006}, see Fig.~\ref{fig:1}(a). 
In the reciprocal space map, see Fig.~\ref{fig:1}(c), rings appear caused by 
the rotational disorder of the Bi(110) domains. The position of the rings 
perfectly matches to the (011), (112) and ($\bar{1}1$0) Bi planes 
expected for the Bi(110) surface, as depicted in Fig.~\ref{fig:1}(e), 
and corresponding to in-plane distances of 3.28\AA, 4.75\AA\ and 4.55\AA.

Annealing the as grown film up to 400~K, results in the repositioning of the Bi 
Bragg peak in the recorded (00) CTR to L=3.3, corresponding to the 
interlayer distance of 3.94\AA\ for Bi(111) \cite{Hofmann2006}, see 
Fig.~\ref{fig:1}(b). The corresponding reciprocal space map is shown in 
Fig.~\ref{fig:1}(d). For this film, also ring structures in the diffracted 
intensity appear, with the most intense ring position matching the 
(0$\bar{1}$1) 
plane, see Fig.~\ref{fig:1}(f), and corresponding to an in-plane distance of 
4.55\AA. Analogue to the Bi(110) film, the presence of the rings results from 
rotational disordered Bi(111) domains on the  surface. 
The Bi(111) domains show 
slight preferential alignment with respect to the six-fold symmetric substrate 
as can be seen from the increased intensity on the ring close to the (11) CTR, 
see Fig.~\ref{fig:1}(d).

\begin{figure}
\includegraphics[width=0.49\textwidth]{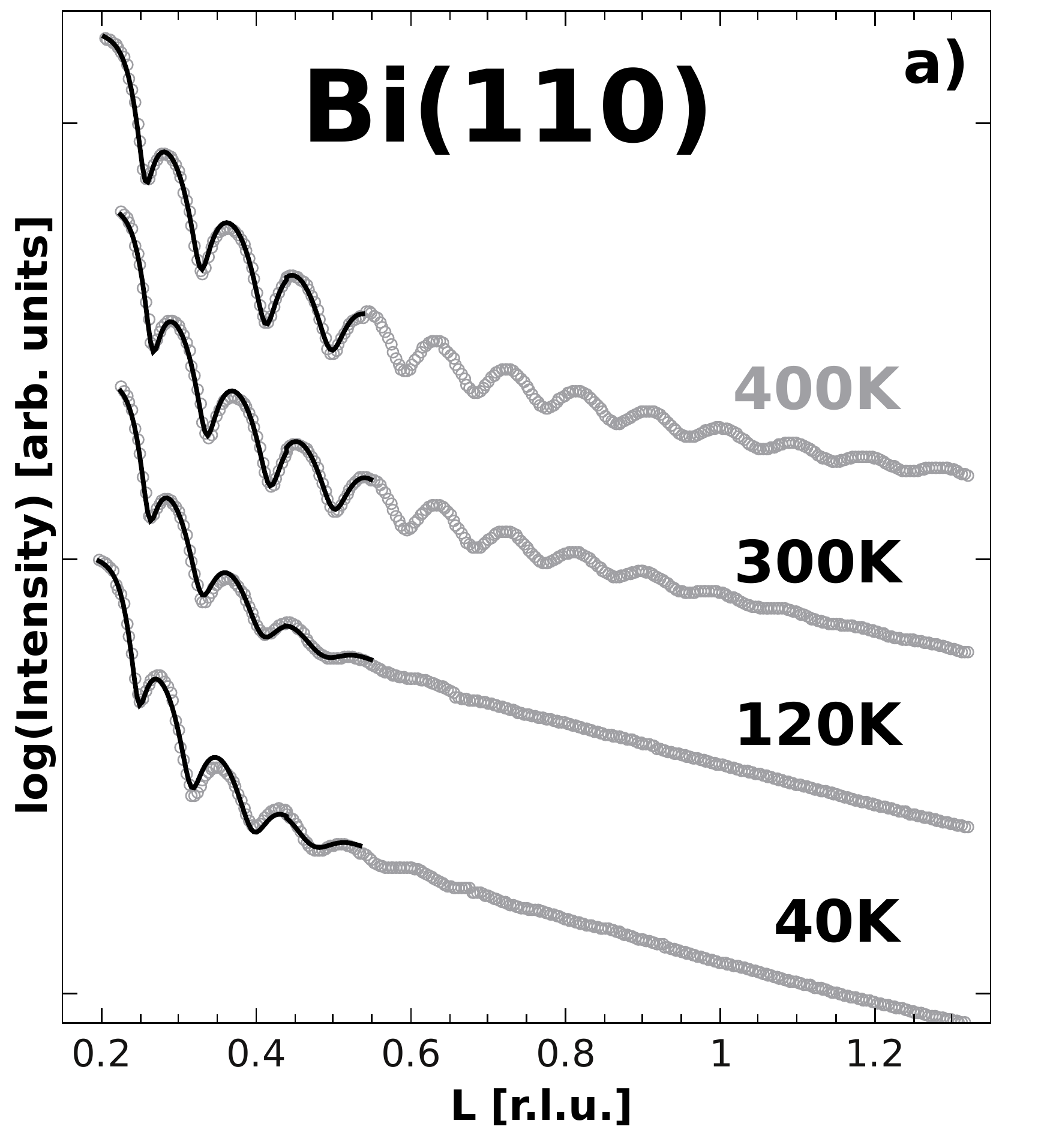}
\includegraphics[width=0.49\textwidth]{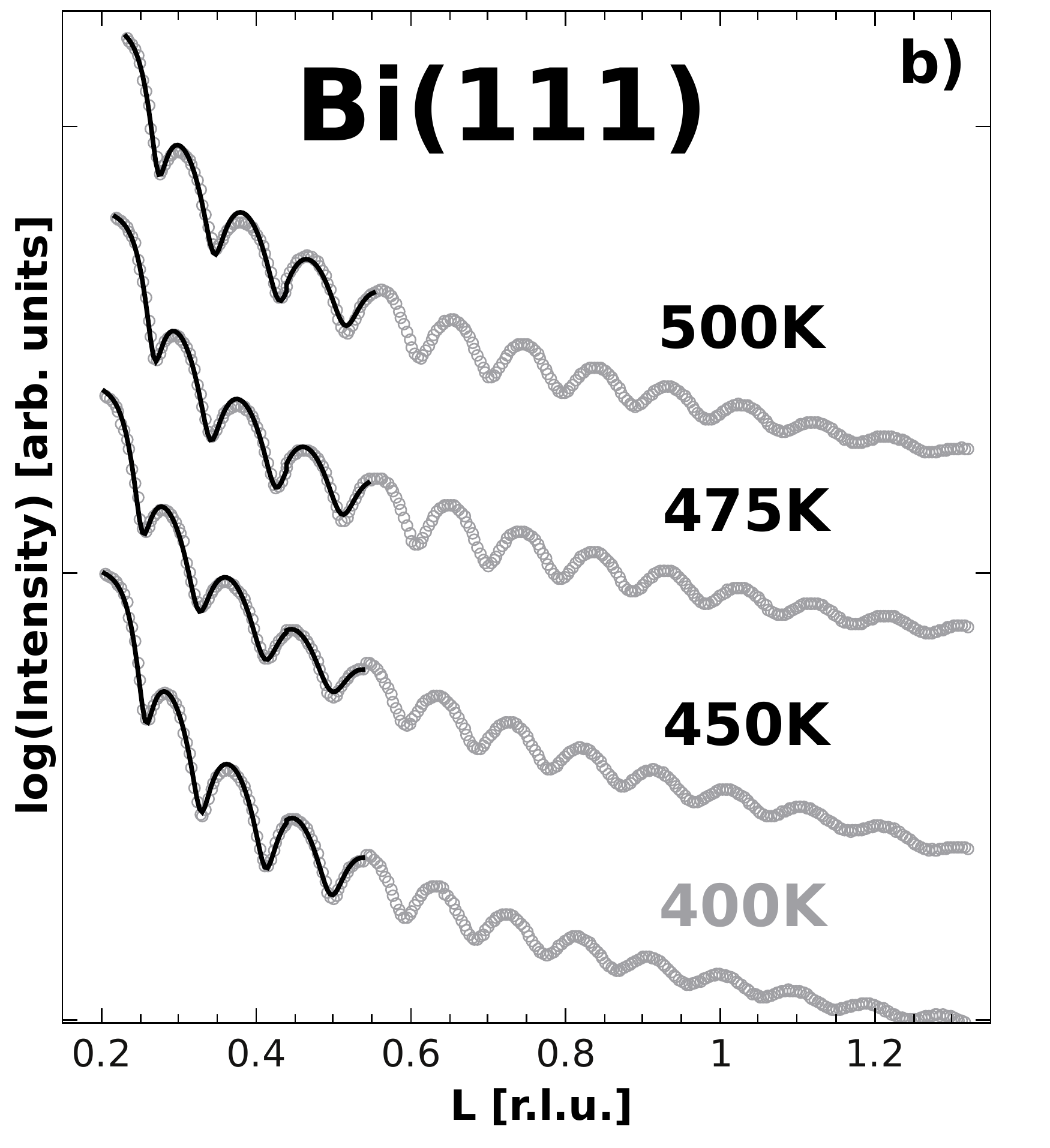}
\caption{(a) XRR scan for a 14~nm thick Bi(110) film grown at 40~K, heated to 
and subsequently measured at 120~K, 300~K and 
400~K. At 400~K, reflectivity scans reveal a Bi(110) and Bi(111) peak, 
indicating the crystallographic orientation transition (see also 
Fig.~\ref{fig:3}(a)). (b) XRR scans for the resulting 14~nm thick Bi(111) film 
heated to and subsequently measured at 450~K, 475~K and 500~K. The solid curves 
in (a) and (b) have been obtained by fitting as described in the text.
}
\label{fig:2}
\end{figure}

\begin{table}
\begin{tabular}{c|c|c|c|c|c}
T(K) & Bi film structure & $R_{rms}^{film}$(\AA) & 
$R_{rms}^{substrate}$(\AA) & $d$(\AA)  \\
\hline
40 & (110) & 7.2 & 8.1 & 142 \\
120 & (110) & 5.5 & 9.1 & 143 \\
300 & (110) & 4.4 & 3.3 & 140 \\
400 & (110) \& (111) & 4.8 & 3.2 & 138 \\
450 & (111) & $<$1 & 5.7 & 136 \\ 
475 & (111) & $<$1 & 3.2 & 137 \\
500 & (111) & $<$1 & 2.6 & 137 \\
\end{tabular} 
\caption{Obtained roughness for the film-vacuum interface 
($R_{rms}^{film}$), the substrate-film interface ($R_{rms}^{substrate}$) 
and the film thickness ($d$) from fitting \cite{GenX} the experimental data as 
shown in Fig.~\ref{fig:2}(a-b). For temperatures $<$400~K, the film grown at 
40~K is pure Bi(110), for temperatures above pure Bi(111).}
\label{tab:1}
\end{table}

\begin{figure}
\includegraphics[width=0.49\textwidth]{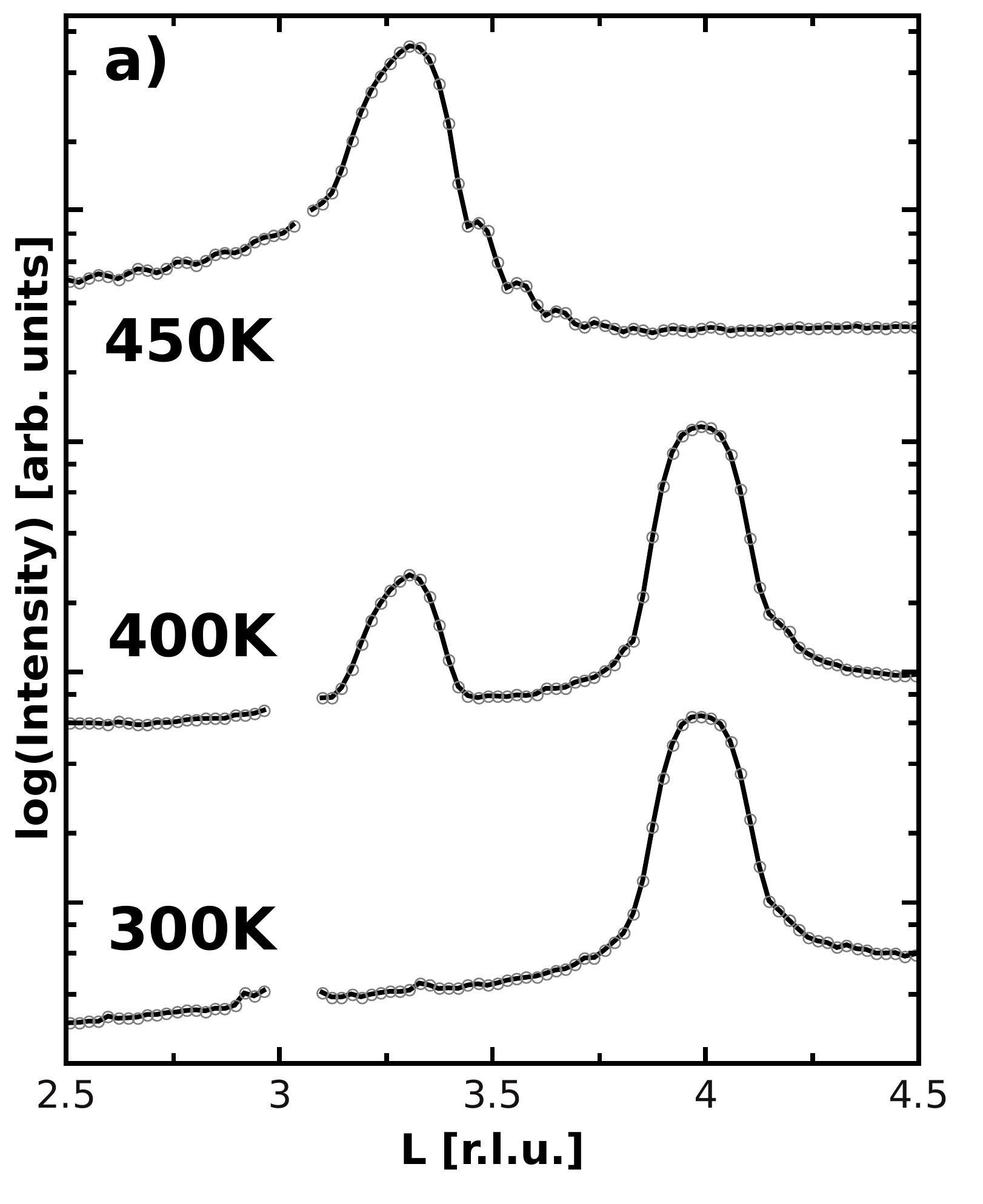}
\includegraphics[width=0.49\textwidth]{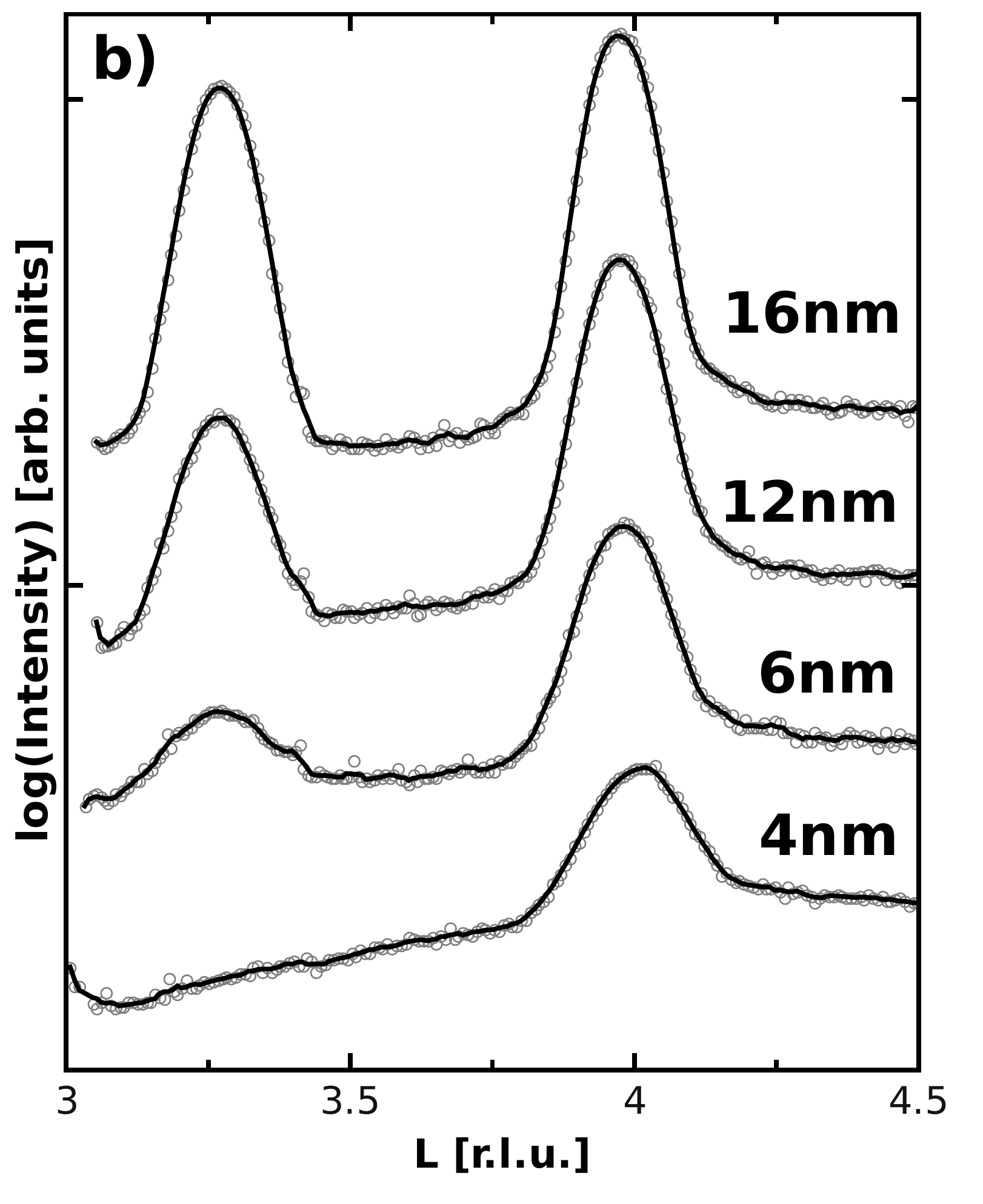}
\caption{a) XRR curves for a 
14~nm Bi(110) film grown at 40~K (corresponding to Fig.~\ref{fig:2}(a)) 
measured 
at 300~K, 400~K and 450~K. At 400~K, the onset of the crystallographic 
orientation transition towards 
a  Bi(111) film is seen where 
at 450~K the entire film is transformed.
(b) XRR scan for increasing Bi film thickness, grown at 
RT. The first 4~nm, Bi(110) domains are grown, followed by Bi(111) domains 
starting around 6~nm total Bi film thickness. 
}
\label{fig:3}
\end{figure}

To test both the Bi(110) and Bi(111) film for their thermal stability and to
investigate the effects of kinetics on the film roughness, we deposited 14~nm 
of Bi on the sapphire surface at 40~K and gradually increased the temperature. 
Fig.~\ref{fig:2}(a) shows XRR curves, revealing the effect of increasing the 
temperature. At higher temperature, the number of 
Kiessig fringes and their amplitude, arising from the constructive interference 
between the X-rays  reflected from the film-vacuum and substrate-film 
interface, increases, indicating the decrease in roughness on both interfaces. 
At 400~K, the Kiessig fringes are also visible beyond L=1.3. The film consists 
at this temperature of Bi(110) and Bi(111) domains, discussed below. The reason 
for the Bi(110) films not transforming below 400~K are the slowed kinetics 
\cite{Campbell1997}. 
When heating the same film beyond 400~K, the film shows a pure Bi(111) crystal 
structure of which the number of Kiessig fringes increase even beyond the film 
Bragg peak at L=3.3, see also Fig.~\ref{fig:3}(a), for temperatures at 500~K and 
above (but below the film melting temperature of $\sim$545~K depending on film 
thickness \cite{Takagi1954}).

In order to quantify the roughness for the film-vacuum and 
substrate-film interface, we model the system as a film of uniform (electronic) 
density on top of a uniform (electronic) dense substrate. Fitting was done by 
using the fitting parameters film thickness ($d$), film-vacuum interface 
roughness ($R_{rms}^{film}$), substrate-film interface roughness 
($R_{rms}^{substrate}$) (see Tab.~\ref{tab:1}),
a background 
resulting from scattering and a 
normalization factor \cite{GenX}. Note that we fit up to 
limited L (here 0.6) to ensure the dynamical 
scattering theory is applicable and stay far from the kinetical 
scattering regime \cite{Vlieg2012}. 

The roughness for the Bi(110) film-vacuum interface can be reduced by about 40\% 
to 4.4\AA\ by heating the sample to RT, see Tab.~\ref{tab:1}. The 
$R_{rms}^{substrate}$ can also be 
greatly reduced, which may be indicative of the rough initial growth due to the 
lattice mismatch described above, resulting in an electronic gradient in the 
profile going from substrate to film. Upon heating, these lattice defects might 
be restored and the film might be (more) decoupled from its substrate as the 
 roughness for the Bi(110) film at RT is similar to Bi(111) films. 
The ultrasmooth Bi(111) films, having a $R_{rms}^{film}$ below 1\AA, 
also reveal 
a decreasing $R_{rms}^{substrate}$ upon increasing temperature. Note, 
that the 
used modeling only includes a fixed and homogenous electronic density value for 
vacuum, film and substrate, giving a very reasonable fit as shown by the solid 
curves in Fig.~\ref{fig:2}(a) and (b). This means that the electronic density 
profile is close to a step function, indicating the decoupling of electronic 
density between substrate and film.

A striking feature in the growth of Bi on sapphire is the appearance of 
the Kiessig fringes when the film is deposited at 40~K, in contrast to 
films deposited at RT and above. We expect this to result from 2D island 
growth as described by Campbell for the growth of 
metals on oxide surfaces \cite{Campbell1997}. According to this model, due 
to the low temperature, the kinetic limitations cause a high nucleation 
density resulting initially in 2D island growth.
Subsequently, the deposited material grows on top of these islands in a 
layer-by-layer fashion, as between the islands the filling proceeds rather 
slowly.
In literature, there are multiple examples of such growth reported 
\cite{Henry1991, Ernst1993, Campbell1994, Baeumer1995}, e.g., continuous Ag 
films on 
ZnO(0001) are demonstrated to grow at reduced temperature 
\cite{Zhang2009a}. For the initial low temperature growth of Bi on 
quasi-crystal 
surfaces, small 2D island formation is reported, transforming towards 
continuous 
films at higher coverages \cite{Bobaru2012}.

Bi(110) films grown at low temperatures can be transformed 
to ultrasmooth Bi(111) films upon annealing to 450~K. However, 
interesting is the region in between, as the film shows a crystallographic 
orientation transition from 
Bi(110) towards Bi(111) in a temperature window of 300 to 450~K, see 
Fig.~\ref{fig:3}(a). The 14~nm Bi film shown in Fig.~\ref{fig:3}(a) is
grown at 40~K and shows only the Bi(110) Bragg peak, heating it to 400~K 
reveals the onset of a Bi(111) Bragg peak. At a temperature of 450~K the entire 
film has transformed into an ultrasmooth Bi(111) as the Bi(110) Bragg peak has 
vanished. The thin film roughness has been reduced (to a 
$R_{rms}^{film}<$1~\AA ) as can be seen from the Kiessig fringes appearing 
around 
the Bi(111) Bragg peak at L=3.3. 
We anticipate this crystallographic orientation transition to be 
resulting from enhanced 
kinetics due to surface pre-melting of the thin Bi film 
\cite{Yaginuma2003}. For flat ultrathin Bi(111) films on Si(111)-7$\times$7 
surface pre-melting occurs at about 350~K \cite{Yaginuma2003}, very similar 
to our observations. 
From this data, it is however not evident how the crystallographic orientation 
transition proceeds. 
One interpretation could be that at 400~K both Bi(110) and Bi(111) domains are 
in competition. A more unlikely interpretation could be that a Bi(111) film 
could be stacked on top of the initially grown Bi(110) film, which would be 
energetically highly unfavorable.
The crystallographic orientation transition of Bi(110) to Bi(111) 
at a critical film thickness is 
subject to ongoing debate in literature. According to Nagao \emph{et 
al.}~\cite{Nagao2004, Nagao2005}, at low film thickness the 
(puckered-layer) Bi(110) is more stable as a result of surface 
effects. As the thickness approaches a critical few layers, the surface effects 
become less dominant, transforming the film to Bi(111), as it 
becomes energetically more favorable. Similar observations were 
done by Bobaru \emph{et. al}~\cite{Bobaru2012} reporting the coexistence of the 
Bi(110) and Bi(111) domains grown at low temperatures and coverage, as well 
as the transformation of Bi(110) to Bi(111) domains at higher 
coverages. There, the coexistence of both crystallographic orientations was 
attributed to the minor 
difference in surface free energy of ultrathin Bi(110) and Bi(111) films and 
Bi(111) films were observed to be kinetically limited at low temperatures. 
Here, we observe solely the growth of Bi(110) at low temperatures, transforming 
to Bi(111) around about 400 K. Surprisingly, the Bi(110) (domains) can be grown 
up to thicknesses of 14~nm, well beyond the critical thickness 
reported by both, Nagao \cite{Nagao2004, Nagao2005} and Bobaru 
\cite{Bobaru2012}.


To test the hypothesis of Bi(110) and Bi(111) domain competition versus 
stacking, 
we 
study films grown around the transition temperature. In Fig.~\ref{fig:3}(b) we 
show the growth of a 16~nm thick film grown at RT where we measured a 
(00) CTR at several different thicknesses. The very thin 4~nm Bi film 
reveals a Bragg reflection for Bi(110) at L=4, but upon increasing the film 
thickness, the Bragg reflection for Bi(111) at L=3.3 starts developing, where 
for thicker films the ratio between both peaks gets more similar. 
Note that the center of the Bi(110) peak for the 4~nm thick film grown at RT, 
is slightly shifted towards higher L as compared to thicker films. This peak 
position corresponds to a slightly compressed average interlayer distance of 
3.24\AA\ as compared to the bulk interlayer distance of 3.27\AA\ \cite{Sun2006, 
Hofmann2006}. From a quantitative Low Energy Electron Diffraction (LEED) 
analysis described in 
literature\cite{Sun2006}, contracted interlayer relaxations are present within 
the first 4 layers of Bi(110), which can heavily contribute to the average 
interlayer spacing on films of only several double bilayers as is the case 
here. 
Although one might anticipate that at RT the Bi(110) film only grows up to a 
certain thickness and then simply transforms into a Bi(111) film, this 
can not be concluded from the increasing area for the peak at L=4 for 
increasing film thickness,
revealing that 
the Bi(110) film continues to grow up to a significant thickness indicative for 
the competition between domains \cite{Bobaru2012}. 
%
Realspace in-situ scanning probe microscopy (SPM) images might help to study 
this initial thin film growth around the transition temperature.

\section{Conclusions}
In summary, we have presented SXRD, CTR and XRR measurements demonstrating the 
controlled growth of Bi(110) and Bi(111) on an atomically well defined 
insulating $\alpha$-Al$_2$O$_3$(0001) substrate. At temperatures as low as 40~K, 
the kinetics of the film growth can be slowed down, resulting in high quality 
pseudo-cubic Bi(110) films, having rotational disordered domains and growing 
solely Bi(110) up to unanticipated thicknesses of tens of nanometers. Bringing 
the film to RT decreases the film-vacuum and film-substrate  
roughness indicative for (electronic) decoupling of the film from the substrate. 
By heating the Bi(110) film above 400~K a crystallographic orientation 
transition occurs to Bi(111).

High quality and ultrasmooth Bi(111) films can be produced by heating Bi(110) 
to 450~K onwards, where the roughness of the film-vacuum interface 
is below 1\AA\ and the roughness between film and substrate 
decreases with increasing temperature. The films show a slight preferential 
alignment with respect to the substrate. 

At temperatures around the crystallographic orientation transition 
($\approx$400~K), the growth of Bi(110) and Bi(111) domains are in competition. 
A film grown at this temperature results in the growth of thin Bi(110) domains 
followed by thicker Bi(111) domains.

The growth of Bi(110) structures on $\alpha$-Al$_2$O$_3$(0001) is unanticipated 
but will have interesting electronic properties\cite{Hofmann2006, Koroteev2008, 
Miao2015}. The growth and possible coexistence of both Bi(110) and Bi(111) films 
on an insulating substrate is very attractive for future electronic and 
practical applications, as the interface between substrate and film, expected to 
reveal topological states, will be protected from influencing oxidation effects 
upon ambient exposure \cite{Tabor2011}. The electronic properties of 
the buried interface could, e.g., be probed by second-order nonlinear optical 
spectroscopy \cite{Hsieh2011}.

\acknowledgments
MJ and TRJB would like to thank Helena Isern and Thomas Dufrane for their
technical assistance. This work is part of the research programme of the 
Foundation for Fundamental Research on Matter (FOM), which is part of the 
Netherlands Organisation for Scientific Research (NWO). 

\bibliography{refs}

\pagebreak
\widetext
\begin{center}
\textbf{\large Supplemental Materials: Controlling the growth of Bi(110) and 
Bi(111) films on an 
insulating substrate}
\end{center}
\setcounter{equation}{0}
\setcounter{figure}{0}
\setcounter{table}{0}
\setcounter{page}{1}
\makeatletter
\renewcommand{\theequation}{S\arabic{equation}}
\renewcommand{\thefigure}{S\arabic{figure}}
\renewcommand{\bibnumfmt}[1]{[S#1]}
\renewcommand{\citenumfont}[1]{S#1}

The Supplemental Material contains Tapping Mode Atomic Force Microscopy 
(TM-AFM), X-ray Photoelectron Spectroscopy (XPS) and Auger Electron Spectroscopy 
(AES) data of the prepared 
$\alpha$-Al$_2$O$_3$(0001) substrate.

\begin{figure}[h]
\includegraphics[width=0.66\textwidth]{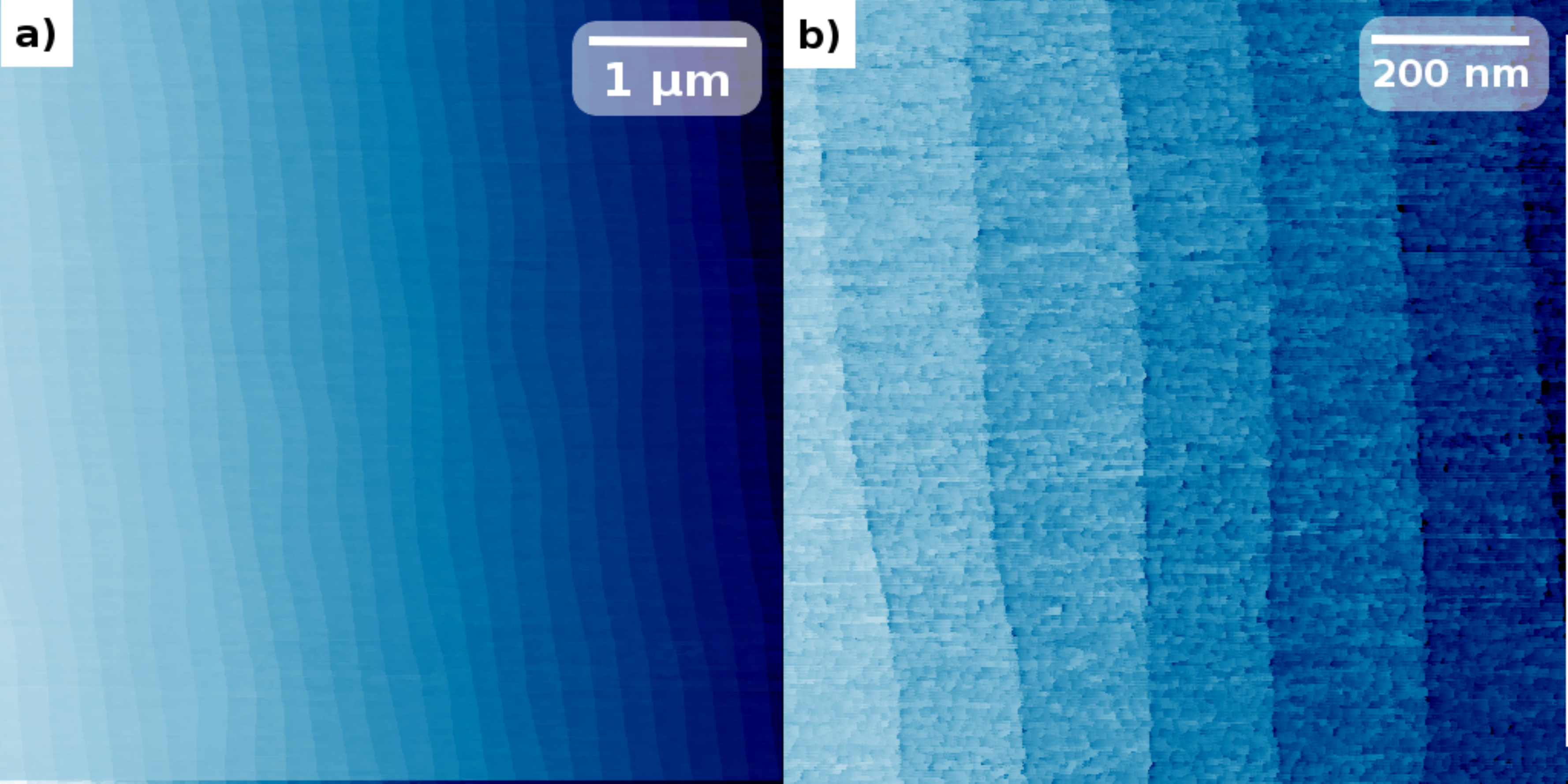}
\includegraphics[width=0.33\textwidth]{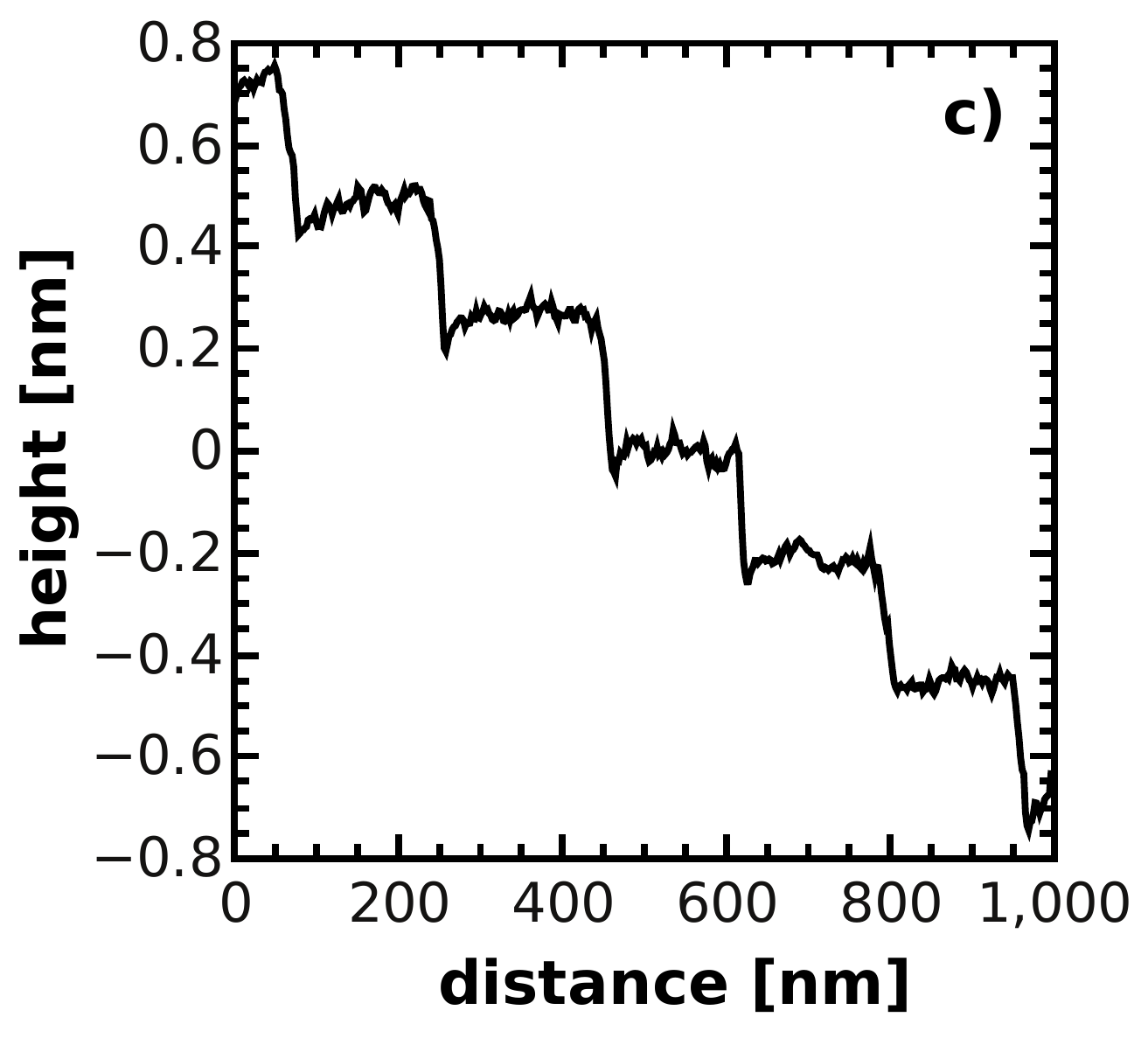}
\caption{a-b) TM-AFM topography of the ex-situ cleaned sapphire surface. c) 
Line-profile across the steps in Fig.~\ref{fig:afm}(b) revealing the substrate 
steps of 0.21~nm.}
\label{fig:afm}
\end{figure}
\begin{figure}[h]
\includegraphics[width=\textwidth]{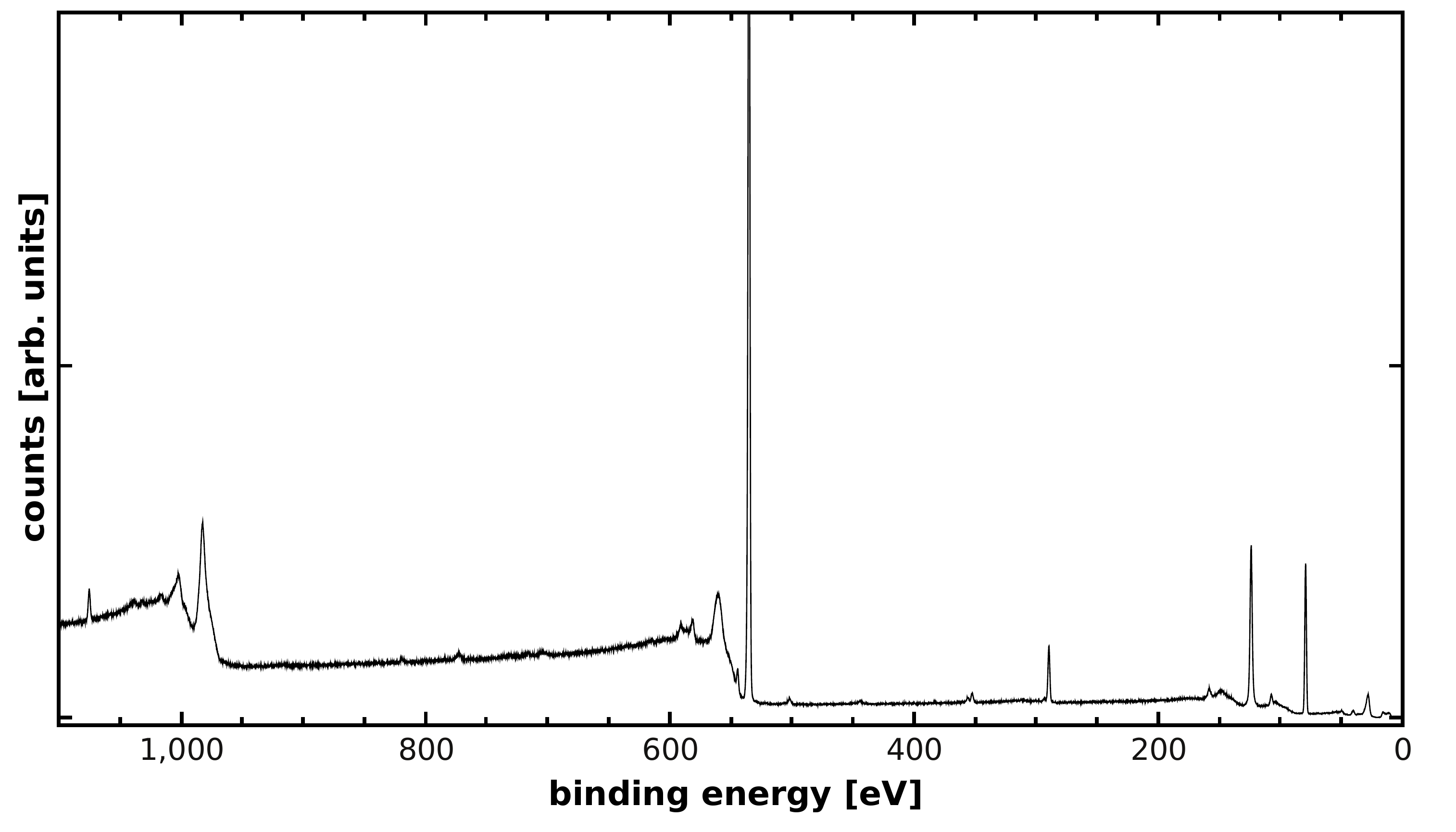}
\caption{XPS curve of the ex-situ cleaned sapphire 
surface.}
\label{fig:xps}
\end{figure}
\begin{figure}[h]
\includegraphics[width=\textwidth]{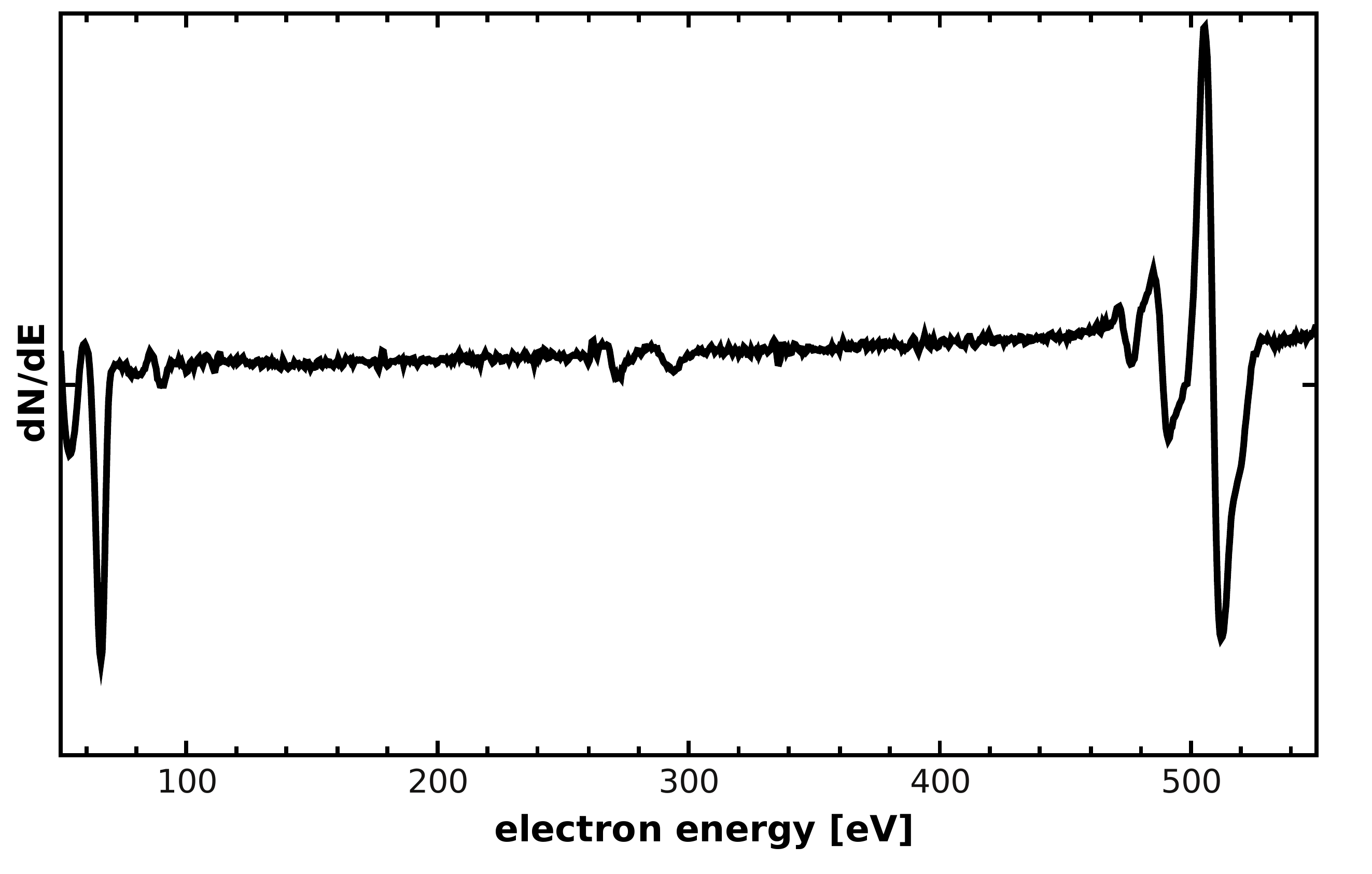}
\caption{AES curve of the in-situ cleaned 
sapphire surface after several sputter/anneal cycles.}
\label{fig:aes}
\end{figure}

\end{document}